# The Solar Optical Telescope for the *Hinode* Mission: An Overview

S. Tsuneta, K. Ichimoto, Y. Katsukawa, S. Nagata[1], M. Otsubo[2], T. Shimizu[3], Y. Suematsu, M. Nakagiri, M. Noguchi

*National Astronomical Observatory of Japan, Mitaka, Tokyo 181-8588, Japan*

T. Tarbell, A. Title, R. Shine, W. Rosenberg, C. Hoffmann, B. Jurcevich, G. Kushner, M. Levay

*Lockheed Martin Solar and Astrophysics Laboratory, B/252, 3251 Hanover Street,*

*Palo Alto, CA 94304, USA*

B. Lites, D. Elmore

*High Altitude Observatory, NCAR, P.O. Box 3000, Boulder, CO 80307-3000, U.S.A.*

T. Matsushita, N. Kawaguchi, H. Saito, I. Mikami

*Communication Systems Center, Mitsubishi Electric Corp., Amagasaki,*

*Hyogo 661-8661, Japan*

L. D. Hill, J. K. Owens

*Space Science Office, VP62, NASA Marshall Space Flight Center, Huntsville,*

*AL 35812, USA*

---

[1]Current affiliation: Kwasan and Hida Observatories, Kyoto University, Yamashina, Kyoto, 607-8471, Japan
[2]Former NAOJ staff scientist
[3]Current affiliation: Institute of Space and Astronautical Science, Japan Aerospace Exploration Agency, Sagamihara, Kanagawa 229-8510, Japan

**Abstract**: The Solar Optical Telescope (SOT) aboard the *Hinode* satellite (formerly called Solar-B) consists of the Optical Telescope Assembly (OTA) and the Focal Plane Package (FPP). The OTA is a 50 cm diffraction-limited Gregorian telescope, and the FPP includes the narrow-band (NFI) and wide-band (BFI) filtergraphs, plus the Stokes spectro-polarimeter (SP). SOT provides unprecedented high resolution photometric and vector magnetic images of the photosphere and chromosphere with a very stable point spread function, and is equipped with an image stabilization system with performance better than 0.01 arcsec rms. Together with the other two instruments on *Hinode* (the X-Ray Telescope (XRT) and the EUV Imaging Spectrometer (EIS), SOT is poised to address many fundamental questions about solar magneto-hydrodynamics. Note that this is an overview, and the details of the instrument are presented in a series of companion papers.

## 1. Introduction

The Sun has strong magnetic fields, and emits intense X-rays from its outer atmosphere. Though observations with the *Yohkoh* satellite point to magnetic reconnection as a necessary ingredient for sporadic coronal heating on various scales from major flares to ubiquitous tiny bursts (Tsuneta, 1996; Yoshida and Tsuneta, 1996), the specific mechanisms of coronal and chromospheric heating remain essentially unknown. Recent progress of the groundbased observations show that the solar magnetic fields consist of an ensemble of fine ($\approx 0".1-0".2$) scale magnetic fields in addition to sunspots and pores (Solanki, Inhester, and Schüssler, 2006). Detailed properties of solar magnetic fields are, however, still unknown due to limitations of spatial resolution and accuracy of magnetic field measurements.

Solar magnetic fields are believed to arise as a result of a global dynamo operating at the base of the convection zone, and also possibly by a local dynamo process (Cattaneo, 1999). Ultimately, we need to improve our knowledge of the solar interior to fully understand the dynamo mechanisms. Even so, the emergence, dispersal and decay of magnetic features at and above the solar photosphere provide an extremely valuable tool for exploring the mechanism of how magnetic flux is generated in the interior and is transported to the surface (Fisher *et al.*, 2000).

The main objective of the *Solar-B* (named *Hinode* after launch, Kosugi *et al.*, 2007, Figure 1) mission is to use a systems approach to understand the generation, transport, and ultimate dissipation of solar magnetic fields with a complex of three coordinated telescopes. For this purpose, *Hinode* carries the X-ray Telescope (XRT: Golub *et al.*, 2007; Kano *et al.*, 2007), the EUV Imaging Spectrometer (EIS: Culhane *et al.*, 2007), and the Solar Optical Telescope (SOT). The energy release and dissipation phase of the magnetic fields are observed with the XRT and EIS, while the SOT performs high-resolution photometric and magnetic observations of the magnetic flux emergence and their subsequent evolution in the photosphere and chromosphere. The uniqueness of the *Hinode* mission is to realize the coordinated and simultaneous observations of the photosphere, the chromosphere, the transition region and the corona to understand how the changing photospheric and chromospheric magnetic fields result in the dynamic response of the coronal plasma.

In the early concept design phase of 1995-1996, the baseline configuration of the SOT was established to be a 50cm diffraction-limited (0".2−0".3) telescope with both a filtergraph and a spectro-polarimeter, considering the balance between the scientific advantage over existing ground-based observations and technical constraints. The filtergraph was needed for high spatial and temporal resolution of the photometric and magnetic observations for both the photosphere and the chromosphere, while the spectro-polarimeter was needed for precise observations of vector magnetic fields. In the course of the development over the 10 years, the progress of high-resolution groundbased observations has been remarkable: Swedish Solar Telescope (SST, *e.g.* Scharmer *et al.*, 2000) delivered ≈0".1 photometric images and ≈0".2 longitudinal magnetograms. Spectropolarimetric observations with the German Vacuum Tower Telescope (VTT, *e.g.* Bello Gonzalez *et al.*, 2005) and the Dunn Solar Telescope (DST, *e.g.* Lites (1996)) reached ≈0".4−0".6 resolution. Indeed, the spatial resolution of the SST may be higher than that of SOT, and the spectro-polarimetric resolution of the VTT and DST may be close to that of the SOT. Nevertheless, simultaneous photometric (imaging) and spectro-polarimetric observations over extended periods of time (> days) with a stable point spread function are critically important for almost all the areas in solar studies. We stress the scientific importance of the uninterrupted observations (shown in movies) for understanding the ever-changing photospheric and chromospheric phenomena.

The 50-cm diameter SOT can obtain a continuous, seeing-free series of diffraction-limited images (0.2−0.3 arcsec) with fully-calibrated high polarimetric sensitivity in the images with a broad spectral resolution (≈0.8nm) in six wavelength bands at the highest resolution. The Narrowband Filter Imager (NFI) provides intensity, Doppler-shift and vector-polarimetric imaging with moderate spectral resolution (≈10pm) in nine spectral lines. When combined, the BFI and NFI observations cover the region from the low photosphere through the chromosphere. The Spectro-polarimeter (SP) provides the line profiles in all Stokes parameters, with the high spectral resolution of 2.15 pm in two magnetically sensitive lines at 630.2 nm. For a typical exposure time, the sensitivity of the SP is 1−5 G in the longitudinal direction and 30−50 G in the transverse direction.

The time cadence ranges from tens of seconds for both photometric images and vector magnetograms in selected NFI lines to a few hours for wide-field scan with SP. The maximum field of view for NFI is 328"×164" with pixel size of 0.08", while that of BFI is 218"×109" with 0.053" per pixel, and SP can view an area of 320"×151" with a pixel size of 0.16" per pixel.

The sun-synchronous orbit of *Hinode* makes possible uninterrupted observations for about 8 months of a year, and is essential for providing a constant heating of the telescope, which is necessary for the opto-thermal stability of the telescope. The downlink of data in nearly every orbit through the ESA Svalbard station significantly contributes to the SOT science by allowing high cadence, a wide field of view, and high resolution observing program.

This paper provides an overview of the Solar Optical Telescope, while the accompanying papers describe the SOT key components in more detail: the Optical Telescope Assembly (Suematsu *et al.*, 2008), the Focal Plane Package (Tarbell *et al.*, 2008), the Image stabilization system (Shimizu *et al.*, 2008), and the instrument polarization calibration (Ichimoto *et al.*, 2008). In Sections 2 and 3, we overview the SOT science and technical system. SOT consists of the Optical Telescope Assembly (OTA) and the Focal Plane Package (FPP), which are described in Section 4. Sections 5 and 7 present the observing modes, control and data flows. Section 6 contains a brief description of the image stabilization system.

## 2. Science Overview

We discuss here some of the outstanding questions to be studied by the SOT and the *Hinode* observatory (Figure 2). Given the excellent quality of the data, we stress that in almost all the research areas with the SOT, interaction with numerical simulations becomes critically important for the quiet sun (Khomenko *et al.*, 2005), for emerging flux (Cheung, Schüssler, and Moreno-Insertis, 2007), for chromospheric waves (Skartlien, Stein, and Nordlund, 2000), and for corona-chromospheric connections (Hansteen *et al.*, 2006; Abbett, 2006; Gudiksen and Nordlund, 2005).

### 2.1 Coronal heating, reconnection and waves: synergy with XRT and EIS

The solar corona is believed to be heated by magnetic reconnection and/or dissipation of MHD waves (Walsh and Ireland, 2003). Direct detection of the various modes of MHD waves with SP is within reach (Ulrich 1996). Very high-frequency MHD waves, if they exist, may play an important role in the heating of active region corona. On the other hand, Parker (1988) proposed that coronal heating is a consequence of reconnection of magnetic fields that have become entangled as a result of motion of their photospheric footpoints. Whether this is true or not should be answered observationally by Lagrangian tracking of individual magnetic elements with the SOT. EIS may detect turbulence associated with jets from the ubiquitous reconnection sites.

*Yohkoh* and *TRACE* (Handy *et al.*, 1999) images show spatially-distinct hot and cool quasi-steady loops, suggesting position-dependent heating rates. Using the HAO/NSO Advanced Stokes Polarimeter, Katsukawa & Tsuneta (2004) found a clear difference in the magnetic filling factor, which is the areal fraction of magnetic atmosphere, at the footpoints of hot and cool loops. SOT allows one to better resolve the specific photospheric conditions (i.e. flows and fields) resulting in hot/cold coronal structures, leading to deeper understanding on the coronal heating.

### 2.2 Active regions and sunspots

The magnetic field in the photosphere is distributed in a very inhomogeneous way,

with sunspots and faculae being the centerpieces of active regions (Solanki, 2003; Weiss, 2006; Ferriz-Mas and Steiner, 2008). There are a number of obvious questions to be pursued with SOT: How are the basic umbral and penumbral structures of sunspots formed and maintained? What drives the Evershed flow in the sunspot penumbral photosphere, and the oppositely-directed inverse Evershed flow in the penumbral chromosphere? How do they disintegrate, and spread their magnetic fragments to the quiet Sun – possibly in a form of moving magnetic features (MMF)? What is the relationship between umbral dots, light bridges and convection? High resolution precise and continuous observations by SOT are contributing uniquely new information about sunspots, moat regions, umbral dots, light bridges, and their sub-surface structures, to name a few.

An important product of SOT's Dopplergram capability is the three-dimensional maps of sub-surface flows and magnetic fields that can be obtained through the application of local helioseismology. This extends our investigation to sub-surface layers, where much of the action is taking place (Sekii 2004; Kosovichev, 2004).

## 2.3 Flux tubes and quiet Sun magnetic fields

A ubiquitous form of magnetic fields at the photospheric level is small-scale, unipolar vertical kilogauss fields, sometimes observed as bright points in the G-band (Berger *et al.*, 1994). Convective collapse (Parker, 1978) may form the kG-strength tubes, and eventually form pores and sunspots, from weaker emerging fields. But, we do not yet know how they are created, evolved and are destroyed. Super-granular diffusion and pole-ward meridional flow was believed to transport fragmented magnetic fields away from sunspots and active regions, and to provide magnetic flux to the quiet Sun (Leighton, 1964). In addition to this, we now know that numerous bipolar ephemeral regions with lifetimes of several hours (Harvey-Angle, 1993) and ubiquitous small-scale horizontal magnetic fields (Lites *et al.*, 1996) with much smaller time-scale emerge and submerge, and as a result, magnetic fields in the quiet Sun are quickly replaced (Title, 2007). Stable long-term observations with the SOT could clarify the demography of these magnetic elements with different origins.

## 2.4 Data-driven simulation of coronal dynamics

Accurate vector-magnetic images obtained with the SOT provide us with time-dependent boundary conditions for coronal magnetic fields. These images allow us to construct 3-D extrapolations of magnetic fields into the corona initially as a snapshot, and eventually a time-dependent evolution is constructed (Welsch *et al.*, 2007). One cautionary note is that we need a tradeoff between polarimetric accuracy and the time required to scan an active region: at least for small-scale flux elements the SP scan duration is usually larger than the time scale for change of a solar feature. If it is successful in reproducing the magnetic field structure with electric current sheets in the corona, the stability analysis as well as the data-driven simulation of the solar coronal dynamics would make possible the forecasting of flares and CMEs in response to the evolution of surface magnetic fields.

## 2.5 Chromospheric heating and dynamics

The chromosphere is maintained by an energy flux ≈10 times greater than that required to maintain the corona (Withbroe and Noyes, 1977). The observational signature for either wave heating (and/or resultant shocks) or magnetic reconnection (or both) should be observed with SOT using both photospheric and chromospheric lines with co-temporal magnetograms (Ulmschneider and Musielak, 2003; Carlsson and Stein, 2004). The chromosphere is highly dynamic, showing ubiquitous jets such as spicules, which can supply mass to the corona and the solar wind. Coordinated SOT and EIS observations studying the thermal evolution of chromospheric ejections are important (Sterling, 2000; De Pontieu, 2004). Furthermore, the chromosphere is closer to the force-free corona, and the chromospheric magnetic fields obtained with the SOT potentially can give better boundary conditions for the coronal field extrapolation than those from the non-force-free photosphere (Metcalf *et al.*, 1995; Leka *et al.*, 2003).

## 3. System Overview

The Solar Optical Telescope is detailed in a series of figures: Figure 3 shows a schematic diagram of the optical systems, Figure 4 shows the electrical configuration, and Figure 5 shows the optical schematic diagram. The OTA and FPP (as are XRT and EIS) are mounted on the satellite optical bench (OBU; Figure 6), which is stable against the launch (mechanical) and orbital (thermal) environment. Figure 7 shows optical interface between OTA and FPP. An accurate alignment of images from three telescopes is crucial, and extensive testing to characterize the thermal deformation of OBU was carried out to ensure the necessary stability. As a result, SOT images are accurately aligned with XRT and EIS images through the observatory-level alignment procedure, which employs successive ladders through nearby images (in terms of wavelength) taken with the different telescopes.

The OTA (Figure 8) consists of the primary mirror, secondary mirror, heat dump mirror (HDM), collimator lens unit (CLU), secondary field stop (2FS), tip-tilt fold mirror (CTM-TM), and the polarization modulator (PMU). Located in OTA, the PMU is controlled by the FPP due to its critical timing with that of CCD exposures needed for polarization modulation. The OTA has two deployment doors for the heat dump window and the entrance aperture, which serves as the entrance pupil. The OTA main structure, to which are mounted these critical optical components, is a precision truss with zero-expansion graphite-cyanate composite material.

The FPP (Figure 9) has a re-imaging lens followed by the beam splitter. The effective combined focal length is 1550 cm (f/31), and the depth of focus in the FPP focal plane is about 400 microns. The focus is adjusted by moving the re-imaging lens through commands from the ground. The re-imaging lens has a stroke of ±25mm in order to have a sufficient margin based on the focus budget breakdown table, which has numerous deterministic and statistical factors.

On the downstream side of the beam splitter are the broadband and narrow-band filter channels sharing a common CCD camera, the spectro-polarimeter, and the correlation tracker. A non-polarizing beam splitter divides the light between SP and the filtergraph, and then the polarizing beam splitter in the filter channel transmits the p-polarized light to the NFI and the s-polarized light to the BFI.

The FPP electrical box (FPP-E) has the computer for controlling the FPP and performs onboard data-processing such as Stokes demodulation. The other electrical box (FPP-PWR) contains the power supply for the entire FPP subsystem. The Mission Data Processor (MDP) controls FPP observations. It follows the observation tables located in MDP, and processes science and housekeeping data from FPP. The CTM-E box has another dedicated computer for the servo control with piezo-driver electronics (CTM-TE box) for driving the tip-tilt mirror (CTM-TM) located inside OTA. The FPP computer and the CTE-M computer directly communicate with each other (i.e. handshake, without any involvement from MDP) to close the control loop for the image stabilization. All the commands from the ground to FPP and CTM-E go through MDP.

The temperatures of the instruments directly affect instrumental safety, and the optical performance for both the OTA and the FPP. Maintaining the instrument temperatures within the desired ranges is one of the critical functions of the system. There are numerous temperature sensors, some of which are fed to the servo controller. OTA has operational heaters to maintain the temperatures of critical optical components, and decontamination heaters to maintain the temperatures of critical optics higher than those of the telescope environment before opening the primary door. The CTM-E controls operational and de-contamination heaters for the OTA. The FPP zone heaters maintain the entire FPP temperatures at 20±1ºC, and also has decontamination heaters for CCD bakeout. The OTA and FPP have survival heaters controlled by the spacecraft heater control electronics in case the primary power for the science instruments is cut off, then the spacecraft will survive.

The observing tables are uploaded from the ground and provide extremely flexible observing sequences for the SOT and the XRT (for details on autonomous XRT observing control with MDP, see Kano *et al.*, 2007). The MDP sends SOT/FPP the macro-commands, which contain all the parameters and instructions to perform the desired observations based on the uploaded tables. The housekeeping data and the image data with header information are separately sent to the MDP from SOT. The image data are compressed, if instructed to do so, combined with the final header information, packetized, and sent to the spacecraft data recorder through the spacecraft central Data Handling Unit (DHU).

The MDP also stores the orbital elements of the spacecraft orbit and the information on the spacecraft pointing, which is used to calculate the Doppler shift of the solar spectral lines. These are sent to the SOT from the MDP, where they are used to compensate in a real time manner the Doppler shift due primarily to the satellite motion in the NFI observations with the tunable filter. Thus, the MDP plays various crucial roles in obtaining smooth and stable SOT observations.

# 4. Optical Telescope Assembly and Focal Plane Package

### 4.1. OTA Optics

The OTA is the diffraction-limited aplanatic Gregorian telescope with a 50 cm aperture primary mirror (Suematsu *et al.*, 2008). Table 1 summarizes the OTA main characteristics. The spatial resolution is specified in terms of the Strehl ratio: the Strehl ratios of the OTA and FPP should be individually better than 0.9 at 500nm, and the combined Strehl is higher than 0.8. Following a conventional definition of the diffraction limit (*e.g.* Maréchal criterion), we set these numbers as a goal. The optical tests simulating in-orbit condition of OTA on the ground (Section 4.3) demonstrated that the OTA had the Strehl ratio better than 0.9 at the wavelength of 500 nm, and the measured FPP Strehl averaged over the field of view is very close to or exceeds 0.9 (Tarbell *et al.*, 2008). The post launch performance appears to meet the goal, and is described in Suematsu *et al.* (2008).

The primary and secondary mirrors are made from ULE. The light-weight (14 kg) primary mirror is supported by an elaborate kinematic mount system to fully meet the stringent requirements on the surface deformation over a wide range of temperatures. At the same time, the fragile primary mirror with its mount system had to survive the severe launch conditions (vibration, acoustic and shock loads) of the ISAS/JAXA M-V solid-booster launch vehicle. The secondary mirror is supported by the fixed invar/titanium mount. Both mirrors are coated with protected silver.

The distance between the primary and secondary mirrors was set at 1.5m, considering the amount of space in the crowded spacecraft, the opto-mechanical tolerance, and the manufacturability of the low f-number primary mirror. The aluminum-made heat dump mirror (HDM) at the primary focus reflects the unused solar light (heat) outside of the 400" dia. field-of-view (FOV) into space through its side window. Since the heat dump mirror is illuminated ×1500 solar, special development and testing efforts were taken for its enhanced silver coating. The moderate temperature (20−40 ºC) of the heat dump mirror is achieved through a high reflectivity and the innovative radiation-cooling design of the mirror.

The conical field-stop is located at the secondary focus to limit the field of view to 361 ×197 arcsec (note that the widest observing field of view is 328×164 arcsec). It is designed so that the light discarded by the secondary field stop will be reflected back via the same route through the secondary and primary mirrors to space.

The OTA has the collimating lens unit (CLU), the polarization modulator (PMU), and the tip-tilt mirror (CTM-TM) behind the secondary focus. The CLU has a short focal length of 37 cm to deliver parallel light to FPP and to create the exit pupil (i.e. the image of the entrance pupil) in the vicinity of the PMU and the CTM-TM. This location of the exit pupil was quite fortunate for the SOT program: in the system-level optical test, we discovered an unacceptable degree of astigmatism – probably caused by the primary mirror. Very late in the program, we decided to add corrective optics (a single cylindrical lens) at the exit pupil to completely remove the astigmatism, which kept the hardware change to minimal.

The CLU consists of six lenses with a IR-rejection filter at its entrance, and is aberration-free (achromatic) and practically instrument-polarization free for the entire range of observing wavelengths: 380−700nm. The first two lenses, which are radiation-robust fused-silica, are used to protect four inner lenses that are more susceptible to radiation, and the spacecraft bus module behind the CLU serves as a backside radiation shield. The CMT-TM has an enhanced silver coating.

An extremely severe positional tolerance of the secondary mirror, with respect to the primary mirror, had been a concern in the design phase. However, the possibility of introducing an adjustment mechanism for the secondary mirror was not a viable choice due to the limited resources in the OTA development program. During the course of the development, the choice to use a 50 cm mirror turned out to be close to the limit in terms of science, technology, cost, and the stringent constraint of construction time. Also, the 50 cm primary mirror is an intended scientific compromise between the requirements for high spatial resolution and a field of view large enough to cover a typical active region.

### 4.2. Polarization Modulation

The polarization modulator (PMU) is located near the exit pupil, and is a continuously rotating waveplate with revolution period $T$ of 1.6 sec in order to provide polarization modulation. The temperature dependence of the PMU retardation is minimized by utilizing two crystals of compensating thermal coefficients of birefringence: quartz and sapphire. The retardation is wavelength-dependent, but is optimized for the 630.2 nm (with a retardation of 1.35 waves) and 517.2 nm (1.85 waves) observations in the sense that Stokes vectors $Q$, $U$ and $V$ have an equally high modulation efficiency of approximately 0.5. The modulation efficiency at other wavelengths is unbalanced among Stokes $Q$, $U$ and $V$.

The polarization states are represented by the Stokes vectors ($I$, $Q$, $U$, $V$). The linear polarization $Q$ and $U$, and the circular polarization $V$ are converted into sinusoidal variations of intensity by the polarizing beam splitters in FPP. Stokes $Q$, $U$ and $V$ are encoded as harmonic variations of intensity at periods proportional to $T/4$, $T/4$, and $T/2$ respectively. The signal vector $Q$ differs in phase from the signal $U$ by 22.5 degrees (relative to the rotational phase of the waveplate). Demodulation of this signal is done by sampling the intensity 16 times per revolution of the PMU waveplate. $I$, $Q$, $U$, and $V$ spectra are then obtained by either adding or subtracting each sample into the four memories corresponding to the four Stokes states in FPP.

The rotating PMU is completely invisible for non-magnetic photometric observations, and any movement in the image due to its minimum residual wedge is removed by the image stabilization system. All optical elements prior to the PMU are rotationally symmetric about the optical axis (except for the secondary mirror supports) in order to minimize instrumental polarization. Note that the folding tip-tilt mirror follows the PMU.

The FPP has a re-imaging lens after receiving the parallel light from the OTA, and it can therefore be regarded as the pupil reducer. In other words, the optical interface between the OTA and the FPP is intended to be afocal, considerably relaxing the positional tolerance of FPP with respect to the OTA. This allows the OTA and FPP to be separately mounted on the OBU as separate independent instruments without any precision requirement (Figure 6).

### 4.3. Optical Testing

In addition to the usual tests for the space flight hardware such as vibration and thermal-vacuum tests, we performed a number of unique tests for the OTA and FPP first separately, and later jointly. Tremendous efforts were expended to plan, develop, and implement these tests, and some major problems – including the discovery of astigmatism – were found as a result of these tests. All of the problems found in the tests were completely analyzed, and corrective actions, sometimes requiring a hardware fix, were thoroughly taken, with the problems declared to be closed only after re-testing demonstrated the desired performance. A lesson we learned is that extensive and complete testing is essential to success of an advanced space-optics instrumentation mission.

The wavefront-error (WFE) measurements are the fundamental test done in the laboratory environment and in the thermal vacuum chamber. A large rotatable precision optical flat was located in front of the OTA entrance aperture to measure the telescope WFE (using a double pass) under auto-collimation. The interferometer is located at the position of the FPP to measure the OTA WFE. The WFE maps reveal a large triangle-astigmatism mainly due to gravity deformation of the primary mirror, but this was cancelled out by adding another WFE map taken with the OTA and the optical flat in the upside-down configuration to cancel the effect of gravity in the OTA. This methodology provides us with means to measure the WFE in a zero-gravity situation. This test was repeated many times during the alignment of critical optical elements, as well as after any mechanical test such as vibration and shock tests.

The Sun test is a unique start-to-finish observation of real sunlight, which was introduced to the clean room through the heliostat by combining the OTA and FPP with the flight electronics (Figures 6). A polarization calibration to obtain the Muller matrix of SOT as a whole (OTA plus FPP) was carried out extensively using the linear and circular polarizers at the entrance aperture of the OTA telescope (Ichimoto *et al.*, 2008).

The opto-thermal test was done to measure the OTA WFE with the large optical flat in the thermal vacuum chamber. Only the interferometer was located outside the chamber in this configuration. A special shroud was prepared to simulate the expected high temperature-gradient along the optical axis while in orbit, and with this setup we were able to verify the OTA optical performance in an environment close to the one for the actual observations. Other system-level (post-assembly level) tests for the OTA include the temperature cycle test, vignetting test, scattered light measurement, focus test, and throughput measurement.

After the delivery of SOT to the spacecraft systems, the SOT (as well as all the other instruments) was integrated to the *Solar-B* spacecraft, and the final series of tests – including vibration and shock tests – continued at ISAS for about one year, at which time instrument builders usually no longer have any access to the instrument to confirm the critical optical performance. However, the SOT is equipped with an optical maintenance port, which is a small hole on the OBU located around the OTA-FPP optical interface (Figure 7). Even after the full installation of SOT to the spacecraft, we were able to measure the WFE of OTA with the optical flat located in front of the OTA aperture and with the interferometer at the port (we have a special optical GSE to introduce light to the OTA-FPP optical interface). The maintenance port was also used to check that the FPP CCD functioned and to confirm the FPP internal and external alignments. Via the optical maintenance port, we repeatedly had opportunity to check the optical health of the OTA and FPP, especially after harsh environment tests, up to delivery of the *Solar-B* spacecraft to the launch site. This was a tremendous help, and gave us strong confidence on the in-orbit performance of all the systems.

### 4.4. Structural and Thermal Properties

*4.4.1 OTA*

The OTA has a precision truss-structure consisting of graphite-cyanate composite-material pipes and honeycomb panels, which were developed especially for the OTA. They have a very low coefficient of thermal expansion (0.05ppm per 1ºC). The main structure is unique in the sense that all the building-blocks are essentially connected with adhesive (not with bolts and pins) to have the required dimensional stability against temperature change and severe mechanical environments (vibration and shock). The OTA had to be light-weight due to the stringent weight budget situation of the spacecraft, but it had to withstand the violent vibration, acoustic, and shock loads imposed by the ISAS/JAXA M-5 launch booster. The total weight of OTA is about 103 kg, and FPP is about 46 kg (without including the separately-located electrical boxes).

The light-weight ULE primary mirror is mounted on the mirror cell (honeycomb plate) through the kinematic mount mechanism, and the CLU, PMU and tip-tilt fold mirror (CMT-TM) are also tightly mounted on the mirror cell. The secondary mirror and HDM are mounted on the spider structure located on the other end of the telescope. The kinematic interface to the cylindrical optical bench unit (OBU) is through the stiff central ring plate located near the mid-point of the telescope's structure. The main structure of the telescope is surrounded by the telescope external housing, which is mounted only on the central ring. The external housing has the entrance aperture (entrance pupil) and heat dump window. Those two apertures have deployment doors to maintain an ultra high level of cleanliness of the telescope during testing and launch, and to prevent an invasion of sunlight into the telescope during the initial outgassing period in orbit.

Although sunlight outside the field of view is reflected into space by the heat dump mirror (HDM), optical elements such as primary mirror illuminated by intense sun light inevitably absorb some fraction of the energy. The solar aborptance is about 6.5% for the primary mirror. All of the optical elements are radiatively coupled to the telescope structure, and the heat dump mirror (HDM) has special fins to dump the absorbed heat. The main internal heat source is the primary mirror located aft of the telescope and the HDM. The heat absorbed by these optical elements is eventually dumped through the large sun-facing radiator (OSR) at the entrance aperture on the sun side, since the backside of the OTA is occupied by the spacecraft bus module, and the OTA is thermally decoupled from the spacecraft. Heat pipe is not used in the telescope system, and the radiation coupling is the primary heat transfer path.

A high temperature-gradient along the Z-axis (from 30ºC at the primary mirror to below 0ºC at the secondary mirror) is needed to transport the heat from the aft to the forward section. The OTA's temperature was determined by designed balance between the solar heat input and the heat dump efficiency from the inside of the OTA, which is regarded as a thermal cavity. Extensive efforts were made to experimentally verify this unique thermal design concept by utilizing two large-scale spacecraft-level thermal vacuum tests (Figure 10). In the thermal balance test of the proto-model OTA and the spacecraft, solar heat input to the individual optical components was simulated accurately by non-flight heaters attached to the optical elements.

There are three different heater systems to maintain the OTA temperatures. Operational heaters maintain the temperature of optical components with special temperature requirements. For example, the temperature of the CLU has to be kept above 25ºC to avoid instrument polarization.

*4.4.2 FPP*

The FFP structure consists of an aluminum honeycomb optical bench with side panels and a cover plate. Note that because of the large depth of focus, using aluminum with large CTE does not pose a problem. The FPP box is mounted on the OBU by a spacecraft-provided kinematic mount. The FPP is thermally isolated from the OBU. Each CCD detector is cooled by its own dedicated radiator. The thermal and structural design of the FPP is described in detail by Tarbell *et al.* (2008).

**4.5. Contamination Control**

A stringent contamination control program has been implemented from the early design phase through testing and the launch to avoid any increase of heat input due to slight darkening of the optical surfaces when contaminated with organic materials. The temperature increase of the primary mirror would result in deformation of the mirror figure due to a difference in CTE between the ULE glass and attached super-invar pads. Therefore, all the flight components are thoroughly baked out and their final outgas rates are quantitatively monitored with the Thermoelectric Quartz Crystal Microbalance. The OTA orbital lifetime (in terms of contamination degradation) is predicted by means of the OTA mathematical contamination model using the measured outgas data.

The primary and secondary mirrors and the HDM have dedicated decontamination heaters that maintain the temperatures of the critical optical components at least 10ºC higher than their surroundings during the high-outgas phase after the launch and during backfill period of the thermal vacuum test. Only the side door was quickly opened to vent the gas from the telescope after launch. The mathematical contamination model was used extensively to predict the outgassing period, after which the telescope main door was opened to introduce sunlight and allow for the start of observations!

Note that the FPP does not necessarily need to have such a stringent plan for contamination control because hazardous solar UV light is essentially absorbed by OTA, and the FPP has a closed structure without any exposed aperture.

# 5. SOT Observing Modes

With the CLU and the tip-tilt fold mirror, the OTA delivers a pointing-stabilized parallel beam to the FPP. The FPP is configured with the re-imaging lens followed by the beam-splitter for the filtergraph, the spectro-polarimeter, and the correlation tracker channels. FPP performs both filter (FG) and spectral (SP) observations at high polarimetric precision, and both types of observation may be performed simultaneously yet independently in response to the macro-commands from MDP.

In filter observation, a 4K×2K CCD camera is shared by the broadband (BFI) and narrowband (NFI) filter imagers, and the NFI and BFI are selected by a common mechanical shutter. The SP and CT have their own CCD detectors. The NFI uses a tunable birefringent filter (Lyot filter) to record filtergrams, Dopplergrams, and longitudinal and vector magnetograms across the spectral range from 517.0−657.0 nm, including several spectral lines: Mg ib (517.3nm; chromospheric Dopplergram and Magnetogram), Fe i (525.0, 524.7, 525.0nm), Fe i (557.6nm), Na i (589.6nm; chromospheric Dopplergram and Magnetogram), Fe i (630.3, 630.2nm), and H i (656.32 nm; Chromospheric structure). The BFI has interference filters to image the photosphere (CN i 388.3 nm, CH i 430.5 nm) and low chromosphere (Ca ii H 396.9 nm), and to make blue (450.5 nm), green (555.1 nm), and red (668.4 nm) continuum measurements for irradiance studies.

The SP is an off-axis Littrow-Echelle spectrograph that records dual-line (Fe i 630.25 nm and Fe i 630.15 nm) dual beam (with the polarization beam splitter, which is a polarization analyzer, in front of SP CCD) Stokes spectra for high precision Stokes polarimetry.

The time sequencing of the science data acquisition by SOT is controlled according to observation tables (one for FG and the other for SP) on the Mission Data Processor (MDP), as will be described in Section 7.

### 5.1. Filter Observations

The BFI (Table 2) produces photometric images with broad spectral resolution in 6 bands (CN band, Ca ii H line, G band, three continuum bands) at the highest spatial resolution available from the SOT (0.0541 arcsec/pixel sampling) and at a rapid cadence (<10 s) over a 218×109 arcsec FOV. Exposure times are typically 0.03−0.8 s, but longer exposures are possible, if desired. The BFI allows accurate measurements of horizontal flows and temperature in the photosphere, and measurements in the ultraviolet bands will permit identification of sites of strong magnetic field. The BFI observes the Ca II H line around line center. These BFI filters have FWMH bandwidth of 0.3−0.7nm, and obtain images not subject to Doppler motion.

The NFI (Table 2) provides intensity, Doppler, and full Stokes polarimetric imaging at high spatial resolution (0.08 arcsec/pixel, somewhat coarser sampling than the BFI) in any one of 10 spectral lines (including Fe lines with a range of sensitivity to the Zeeman effect, Mg ib, NaD lines, and H alpha) over the full field of view (328×164 arcsec). The spectral lines span the photosphere to the lower chromosphere for diagnosis of dynamical behavior of magnetic and velocity fields in the lower atmosphere. The spectral bandwidth of the Lyot filter is ≈95 mA at 630 nm, and the wavelength center is tunable to several positions in a spectral line and its nearby continuum. There is no wavelength shift across the field of view due to the telecentric beam. It is noted that the edges of the full field of view are slightly vignetted due to the limited size of the optical elements of the tunable filter residing in a telecentric beam. The un-vignetted area is 264 arcsec in diameter. Exposure times are typically 0.1−0.4 s, but like the BFI, longer exposures are possible.

Filter observations mainly produce four types of observables: filtergrams, dopplergrams, longitudinal magnetograms, and Stokes *IQUV* images. Filtergrams are snapshot images acquired from a single exposure for mapping the intensity of the solar features. "Broadband" filtergrams are the only observable made by BFI. "Narrowband" filtergrams are obtained by the NFI for all the spectral lines and nearby continuum located in the NFI spectral windows. The shutter open/close operations are always synchronized to the phase of the PMU. Various combinations of frame size and pixel summing mode may be chosen in order to reduce the data volume at the expense of FOV size and/or spatial resolution. The readout times for the full CCD are: 3.4 s at $1\times1$ summing, 1.7 s at $2\times2$ summing, and 0.9 s at $4\times4$ summing. The readout of a smaller window of the CCD (several discrete sizes from 192 to 2048 rows) is possible in the central $2K\times2K$ pixel area for faster cadence as well as for reduced data volume. The time for reconfiguring mechanisms, including wavelength change by filter wheels, is less than ≈2.5 s. Onboard processing is performed in the FPP to make magnetograms, dopplergrams and Stokes parameters, and data compression is done in the MDP as described below.

Dopplergrams are images of the Doppler shift of a spectral line derived from narrowband filtergrams at several wavelengths. The central wavelength is derived from two or four images uniformly spaced through the line. Onboard memory processing is performed in real time in the FPP to calculate sums, differences and ratios of images, which are sent to the ground separately. The data are converted on the ground to a velocity via a lookup table. The best photospheric line for Doppler measurements is Fe i 557.6 (Lande *g*=0). The rms noise is typically 30 m s$^{-1}$ for an observation with four images.

Longitudinal magnetograms give the location, polarity and a crude estimate of flux of the magnetic field components along the line of sight. Onboard processing in the FPP combines multiple narrowband filtergrams into two-image data (numerator and denominator) for reduced telemetry load. The primary lines are Fe i 630.25 and Fe i 525.02 (for photosphere) and Mg i 517.27 (for low chromosphere). Typical magnetograms take ≈20 s for eight images and have an rms noise of ≈$10^{15}$ Mx per pixel.

Stokes *I/Q/U/V* images are made onboard from narrowband filtergrams at eight phases of the polarization modulator for each wavelength setting. Stokes demodulation is done such that there is minimum noise due to the time change of the Stokes *I*. Analysis of *IQUV* images at multiple wavelengths in a spectral line yields vector magnetic field information (*i.e.* vector magnetograms). Shutterless modes with the frame transfer operation of the CCD are used for higher time resolution (1.6−4.8 s) and sensitivity, although the field of view is restricted by a focal plane mask. With 0.1s exposure, 16 images are taken in a revolution of the PMU waveplate. These images are successively added or subtracted in the four slots of the smart memory to create the Stokes *IQUV* images. The modulation frequency is 2 per PMU rotation for *V* and 4 per PMU rotation for *Q* and *U*. Optionally, longer exposures may be used: with 0.4s exposure (1/4 of the PMU rotation), we can measure only *V*; with 0.2s exposure we can measure *Q*, *U* or *V*; and with 0.1s exposure, we can measure all *QUV*.

The processing in smart memory is identical to that for SP (see Section 5.2.). In shutterless mode, the FOV is $5.1 \times 164$ arcsec for 0.08 arcsec pixels, $12.8 \times 164$ arcsec for 0.16 arcsec pixels, and $25.6 \times 164$ arcsec for 0.32 arcsec pixels. Larger FOV may be obtained using successive exposures or longer exposure times (for partial Stokes sets). Stokes *IQUV* parameters also may be measured using the mechanical shutter. The FOV is up to $82 \times 164$ arcsec for 0.08 arcsec pixels and $328 \times 164$ arcsec for 0.16 or 0.32 arcsec pixels. Up to 0.4 s exposures are possible for *V*, and up to 0.2 s for *Q* and *U*. Note that additional noise sources due to the time between frames and cross-talk from Stokes *I* may appear.

### 5.2. Spectral Observations

The Spectro-Polarimeter (Table 3) obtains line profiles of two magnetically sensitive Fe lines at 630.15 and 630.25 nm and the nearby continuum using a $0.16 \times 151$ arcsec slit. Spectra are exposed and read out continuously 16 times per rotation of the PMU, and the raw spectra are added and subtracted onboard in real time to demodulate them, generating Stokes *IQUV* spectral images. Two spectra are simultaneously taken in orthogonal linear polarizations. When combined during the data analysis after downlink, spurious polarization due to any residual image jitter or solar evolution is greatly reduced. The solar image may be stepped across the slit to map a finite area, up to the full 320 arcsec-wide FOV.

The SP is flexible in mapping observing regions, allowing one to perform suitable observations depending on science objectives. The SP only has a few modes of operation: Normal Map, Fast Map, Dynamics, and Deep Magnetogram. The Normal Map mode produces polarimetric accuracy of 0.1% with $0.15 \times 0.16$ arcsec pixels. It takes 83 min to scan a 160 arcs-wide area: enough to cover a moderate-sized active region. By reducing the scanning size, the cadence becomes faster (50 s for mapping of 1.6 arcsec-wide area), which would be useful for studying dynamics of small magnetic features. The Fast Map mode of observation can provide 30-min cadence for a 160 arcsec-wide scan with $030 \times 0.32$ arcsec pixel size. In the Fast Map mode, the Stokes profiles at two slit positions with each integration time of 1.6 s are summed, and 2 pixels along slit are also summed to have polarization accuracy a factor of 1.15 better than 0.1%. The Dynamics mode of observation provides higher cadence (18 s for 1.6 arcsec-wide area) with 0.16 arcsec pixels, although with lower polarimetric accuracy. In Deep Magnetogram mode, photons may be accumulated over many rotations of the PM, as long as the data doesn't overflow the summing registers. This allows one to achieve very high polarization accuracy in very quiet regions, but at the expense of time resolution.

# 6. Image Stabilization System and Micro-vibration

**6.1 Image Stabilization System**

SOT is equipped with an image stabilization system that greatly reduces the degradation of the image resolution and the polarization cross-talk due to the image jitter. (The polarization cross talk is caused by changes in intensity.) The spatial fluctuations are due to jittering of the spacecraft's attitude and drift, some possible wobbling associated with the PMU rotation, and slow drifts caused by opto-thermal deformation of the instrument structure. The stabilization system is essential for obtaining crosstalk-free polarization and magnetic maps. The required stability is the rms of the displacement to less than 0.03 arcsec (Shimizu *et al.*, 2008).

As described in the previous Section, the image stabilization system (Table 5) consists of the detection of the image jitter in the focal plane by a correlation tracker (CT in FPP), the high-speed transfer of the jitter (error) signal to the software-controlled digital servo (CTM-E), an analog driver (CTM-TE) for the piezo devices, and the tip-tilt mirror (CTM-TM in OTA). The correlation tracker obtains a displacement error from correlation tracking of solar granulation. The fold mirror in OTA near the telescope pupil is a piezo-driven tip-tilt mirror controlled by a closed-loop servo. The system minimizes the jitter of the images in the focal plane CCDs in the frequency range lower than 14 Hz (for nominal gain). This relatively low bandwidth is due to the delay-time in the closed loop needed for the CCD readout.

The CT is a high speed (580 Hz) CCD camera used to detect motions of the images in the focal plane by looking at the solar granulation pattern. The displacement of the live images with respect to the reference image, updated in a specified interval (currently 40 s), is calculated by the FPP computer, and the derived jitter signal is fed to the closed-loop controller.

The commercial piezo devices manufactured by Queensgate Instruments, Ltd., were chosen after testing a few candidate devices. An extensive space-qualification program, including a long-term life test at high temperature and in vacuum, was implemented at NAOJ with the help of the manufacturer. Three piezo devices are used for the two-axis control of the mirror so that even if one of devices or drivers fails, the image stabilization could still work, just with a smaller stroke angle. The CTM-E onboard software supports this contingency mode for uninterrupted observations.

The image stabilization system achieves a remarkable stability of 0.007" ($1\sigma$) in orbit. Since the pupil size is reduced by a factor of approximately 16 at the location of the tip-tilt mirror, the stability is partially caused by the large angle amplification factor of the same amount (between the tilt angle of the mirror and the angle on the celestial plane). As a matter of fact, in addition to the image stabilization system, the excellent spacecraft attitude stability, the structural-thermal design of the instrument, and the stable solar heat input to the telescope due to the sun-synchronous orbit all contribute to the exceptional performance of the telescope.

Note that the correlation tracker produces a displacement signal by using the granules seen in the 11×11 arcsec field of view as fiducial points, and therefore the entire SOT field of view tracks the group motion of granules in the specific small area. It is noted that the entire spacecraft is directed to an observing target, and that the satellite pointing is controlled to track the observing region continuously with the speed of the solar differential rotation.

**6.2 Micro-vibration**

Micro-vibrations are excited by various noise sources (linear-force and moment-force), such as instrument mechanisms as well as satellite gyroscopes and momentum wheels. The frequency range of the micro-vibration is much higher than the bandwidth of the SOT image stabilization system. Tremendous efforts were made to characterize this effect by using the flight spacecraft and telescope to cope with the micro-vibration coming to the OTA, some of which has severe resonance with the telescope structure. The accurate measurement of the micro-vibration level was made through ultra-sensitive accelerometers attached or close to the primary and secondary mirrors and by measuring the light beam fluctuation from OTA through the OTA optical test port with a high speed camera (position sensitive detector). The effect of micro-vibration was decreased by the relocation of noise sources to a location, which has lower transfer function to the telescope, less change in moving frequency to avoid resonance, and a structural improvement and operational workaround.

# 7. SOT Observation Control and Data Flows

The observing sequence of the SOT is entirely controlled by the Mission Data Processor (MDP), following the observation tables (Figures 11 and 12) in the MDP. In this sense, the SOT is slave to the MDP (Figure 4). There are two concurrent observation tables: one dedicated for FG observation, and the other for SP observation. Each table contains several lists of commands for acquiring observables, such as filtergrams, Stokes maps, magnetograms, and dopplergrams. These macro-commands, which have all the information needed to perform the intended observations, are issued from MDP (by reading the tables), and the SOT/FPP faithfully takes the observations. The contents of the tables are uploaded from the ground in science observing plans, and the table uploads usually happen every day.

The control table structure for SOT observations is shown in Figure 11. Flags for the SOT mode transition, which are updated by commands, are maintained in the *current control table*. This table contains parameters such as information on the conversion from the coordinate of a flare detected by XRT to FG/SP coordinate. The MDP calculates the Doppler velocity due to solar rotation and the satellite motion, for which the data are maintained in the *Doppler Table.*

The SOT observing time-line that executes the science objectives is made up of *Observation Programs,* which we call *Sequence Tables* (Figure 11). The observation program, which allows nested loop structures, is the main program, and the maximum number stored in MDP is twenty. One observation program consists of the *main routine* and four *subroutines*. An example is shown in Figure 12. The main routine calls one of the four subroutines with the repeat count and time interval for calling the next line (subroutine). The loop count of the main routine may be specified, with zero denoting an infinite loop. A maximum of eight subroutine calls may be included in the main routine. The individual subroutine then calls the *sequence tables* with the repeat count and time interval for calling the next line (sequence). Subroutines may call a maximum eight sequences. The sequence table is the sequential list of macro-commands with the timing for the next command. Instrument commands for engineering and maintenance purposes may also be included in the sequence tables, a feature which is a great help in operation. One sequence table consists of eight command lines, and a maximum of one hundred sequence tables may be used.

Science data (CCD images) are acquired by the FG and SP CCD cameras. Multiple exposures may be taken to generate observables. The generation of observations is processed in the SOT/FPP in real time, and the processed science data are then transferred to the MDP via a high-speed parallel interface. Because of limited telemetry downlink bandwidth, data are compressed in terms of depth (16 to 12 bit compression) and in terms of two-dimensional images (image compression). The MDP assembles CCSDS packets from the compressed data, and sends them to the spacecraft central Data Handling Unit (DHU) for recording in the spacecraft common Data Recorder (DR).

The MDP has eight look-up tables to perform the 16-to-12 bit compression with different compression curves. For image compression of SOT data, two algorithms are available with different compression parameter tables: one is a 12-bit JPEG DCT lossy compression, and the other is 12-bit DPCM lossless compression. According to studies with simulated SOT data, filtergram data may be compressed to ≈3 bits/pixel by the JPEG algorithm and Stokes data to ≈1.5 bits/pixel, at which point noise due to lossy compression is comparable to the photon noise level in the data, although compression ratio is highly dependent upon the nature of the images. The data compression is done by the dedicated 12-bit JPEG gate-array developed for *Hinode* by the Mitsubishi Heavy Industries,

Ltd.

SOT observations are telemetry-bandwidth limited. Thus, in the planning of observations, wise usage of the spacecraft data recorder is needed, considering the allocation of telemetry downlink stations. The capacity of the *Hinode* data recorder is ≈8 Gbits in total, of which ≈5.6 Gbits may be available for SOT data with an operational agreement of ≈70% allocation to SOT, although this allocation can be changed. Approximately 1.7 Gbits of SOT data can be downloaded through the nominal 4 Mbps high-telemetry channel in one ground station pass (assuming a 10 min duration). If fifteen stations are scheduled per day, then SOT can acquire 25.5 Gbits data per day, and the corresponding post-compression average data rate from SOT is ≈300 kbps. An extreme example is to perform a high-rate burst observation, which provides high-cadence observations with wide field of view. The post-compression maximum data rate is ≈1.3 Mbps. When XRT and EIS observations are not solicited, which depends on the science purpose (so-called SOT-dominant mode), the maximum rate can reach ≈1.8 Mbps. In the burst observation mode, it takes about one hour to completely fill the 5.6 Gbits of space in the spacecraft recorder. That data would require about three station passes for complete downlink.

## 8. Conclusions

The Solar Optical Telescope (SOT) aboard *Hinode* is the largest aperture advanced solar telescope ever launched into space. The SOT consists of the Optical Telescope Assembly (OTA) and the Focal Plane Package (FPP), and it obtains high resolution photometric images from the photosphere to the chromosphere and makes highly accurate measurements of the vector magnetic fields with its filtergraph and spectro-polarimeter. The stable cadence, unaffected by spacecraft night (i.e. eclipses) or bad seeing, is particularly effective in obtaining high-quality movies, from which various discoveries are being made.

The in-orbit performance of the SOT is generally excellent, and has met or exceeded all pre-launch expectations for BFI, SP and image stabilization system. However, images from the NFI unfortunately contain artifacts which degrade or obscure the image over part of the field of view. These are caused by air bubbles in the index matching fluid inside the tunable filter. They distort and move when the filter is tuned, and then usually drift toward the edges of the FOV over time. For this reason, NFI observing is usually done in one spectral line at one or a small number of wavelengths for extended periods of time. Rapid switching between lines is not allowed. Software changes made since launch have given us considerable control over the location of the bubbles; targets can usually be placed in large blemish-free areas of the CCD. Tuning schemes have been developed which permit tuning to different positions in a line profile without disturbing the bubbles. This has enabled collection of most of the expected NFI observations. Flat field correction of NFI images is still a challenge, but progress is being made on this; magnetograms and Dopplergrams are usually self-correcting since they are made from ratios of intensity differences. Details on the NFI performance will be published elsewhere.

Notes and Acknowledgements

The Solar Optical Telescope (SOT) aboard *Hinode* is the result of a fruitful international collaboration between Japan and the United States. The SOT design meetings were held sixteen times, either in Tokyo or Palo Alto, until the start of the final level spacecraft testing in August 2004. The extensive week-long interaction in the design meetings resulted in the successful design, fabrication, and joint tests of the mechanical, thermal, electrical, optical, and control/guidance components of the instrument. All of the participants in the program were impressed with the rapid development from had initially what appeared to be ambitious program to the sophisticated state-of-the-art instrument at its completion.

The principal investigator (PI) of SOT is Saku Tsuneta. The U.S. PI for NASA had been Alan Title, and Ted Tarbell succeeded him in the fall of 2005. Needless to say, numerous scientists, engineers and managers in Japan, the United States, and France contributed to the program. We sincerely thank those individuals and the organizations they represent.

SOT consists of Optical Telescope Assembly and Focal Plane Package. The optical telescope assembly (OTA) and the image stabilization sub-system (CTM) of SOT were built by the *Solar-B* project office and the Advanced Technology Center (ATC) of the National Astronomical Observatory of Japan (NAOJ) and their industry partners. The prime industry partner for OTA/CTM is the Communication Systems Center of the Mitsubishi Electric Cooperation, along with the participation of SAGEM REOSC (primary, secondary, heat dump, secondary field stop mirrors), Canon, Inc. (collimator lens unit (CLU) fabrication), Genesia (CLU optical design and astigmatism corrector), Sankyo Optics Industry Co., LTD., and Okamoto Optics Works, Inc. (astigmatism corrector), Systems Engineering Consultants (SEC: CTM flight software), Mitsubishi Space Software (CTM digital electronics and OTA thermal design), Queensgate Instruments, Ltd. (CTM actuators), and Mitsubishi Heavy Industries., Ltd. (MDP). Koichi Waseda of the NAOJ ATC designed and fabricated the flight IR filter for CLU.

FPP consists of a wide-band camera (WB), narrow-band camera (NB), spectro-polarimeter (SP), and correlation tracker (CT). FPP was built by the Lockheed Martin Advanced Technology Center, with participation from the High Altitude Observatory of the National Center for the Atmospheric Research (who are mainly responsible for SP), and NAOJ. Significant subcontractors included E2V and Mullard Space Sciences Laboratory (CCD detectors), Barr Associates and Andover Corporation (filters), Vision Composites (structure), and Horber Magnetics (motors).

We would like to thank Hirohisa Hara, Tomonori Tamura, Naoko Baba and her team at JAXA/ISTA (contamination control), Ryohei Kano and Masahito Kubo (Onboard Doppler correction algorithm), Ken Kobayashi (CTM analog electronics), Kenji Minesugi (structural, JAXA/ISAS), Akira Onishi (thermal, JAXA/ISAS), Keiichi Matsuzaki (MDP). (Personnel without affiliation are from NAOJ at the time of development.) We also thank Yasushi Sakamoto, and Naoki Kohara for their contribution to the testing.

Izumi Mikami, Hideo Saito, Tadashi Matsushita, and Noboru Kawaguchi led the SOT program at the Communication Systems Center of the Mitsubishi Electric Corporation. Lead engineers of the Mitsubishi include Toshitaka Nakaoji (OTA structure), Kazuhiro Nagae (OTA thermal), Yasuhiro Kashiwagi (CTM systems), Osamu Ito (CTM analog electronics), Yoshihiro Hasuyama (integration/inspection/assembly of critical optics), Kazuhide Kodeki (CTM guidance and control), Masaki Tabata (CTM mechanism development), Norimasa Yoshida (guidance and control, microvibration), Tsuyoshi Ozaki (composite material), Nobuaki Kaido (OTA thermal), Shusaku Inoue (CTM digital electronics), and Jun Nakagawa (OTA deployment door).

We also thank Renaud Mercier Ythier , Luc Thepaut, Eric Ruch, and Daniel Mouricaud (SAGEM/REOSC), Hideo Yokota and Masaharu Suzuki (CANON), Masayuki Nagase (SEC), and Kim Streander (HAO) for their superb work.

Tim Gordon and Therese Errigo of the Swales Aerospace contributed to various OTA contamination control issues, with support from the NASA MSFC (Keith Albyn and Larry Hill) and the US Naval Research Laboratory (Clarence Korendyke). Charles Powers of NASA Goddard Space Flight Center provided additional qualification for the DEB dampers of the OTA doors. Jim Bilbro and Scott Smith (NASA MSFC) advised the OTA program in resolving some critical optical issues.

Michael Levay, Bruce Jurcevich, and Chris Hoffmann led the FPP development program in the U.S. Department heads included Bill Rosenberg and Gary Kushner (systems engineering), Chris Hoffmann (assembly, integration and test), Dick Shine (tunable filter), David Elmore (SP), Chris Edwards (electrical, CT, and CCD cameras), Dnyanesh Mathur (software), Barbara Fischer (mechanical), Dave Akin (mechanisms), Ericka Sleight (thermal), and Tom Cruz (logistics). The *Solar-B* project office at NASA Marshall Space Flight Center extensively oversaw the SOT program in the United States. The NASA project office led by Larry Hill consisted of Jerry Owens, Robert Jayroe, Barbara Cobb, Vernon Keller, Danny Johnston, Charlotte Talley, and Spence Glasgow.

Sadanori Shimada and his team working in spacecraft systems (at Kamakura Works of the Mitsubishi Electric Corporation) supported the OTA development and SOT integration to the spacecraft.

S. T. would like to express deep appreciation to the former and the present director generals of the National Astronomical Observatory of Japan, Prof. Norio Kaifu and Prof. Shoken Miyama, for their strong support to the program. The authors thank Gary Kilper for comments on the paper.

Figure Legend

**Figure 1**

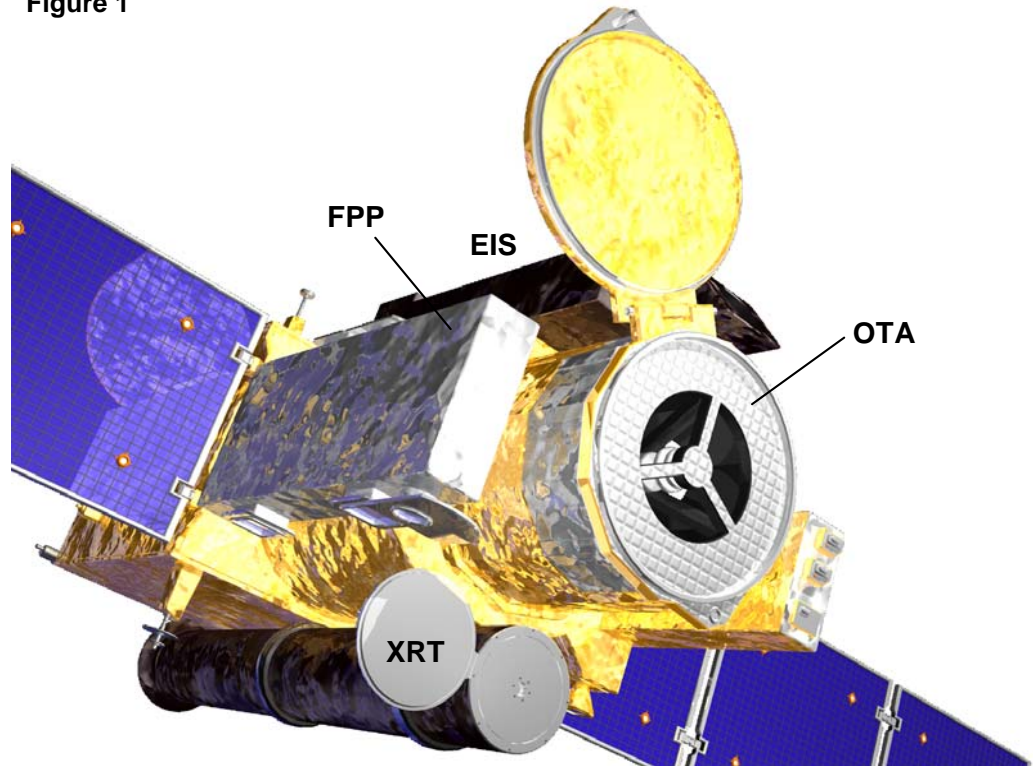

Figure 1

*S*olar-B (*Hinode*) outlook in orbit

**Figure 2**

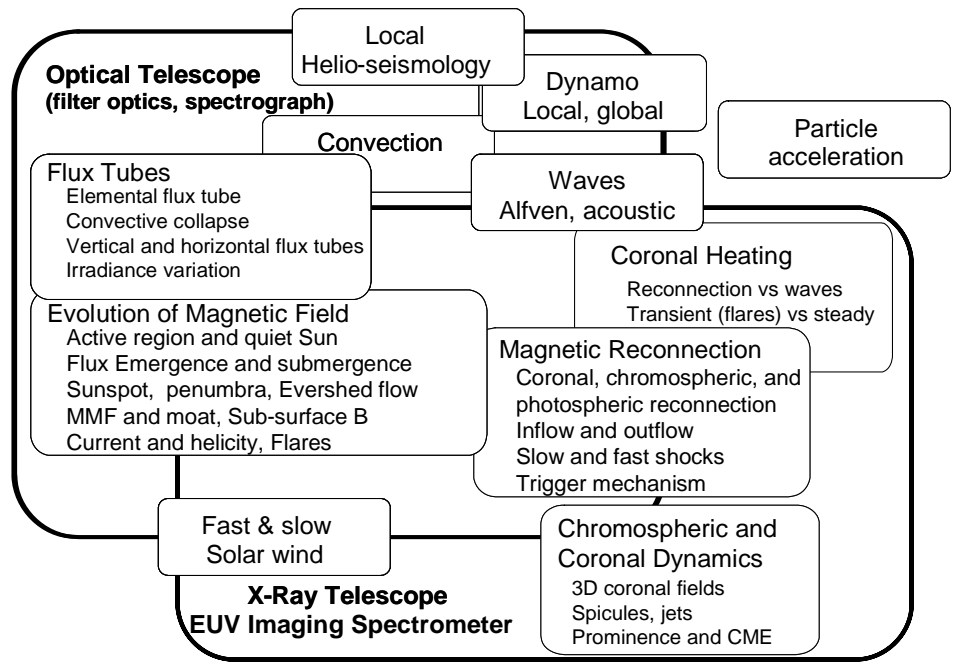

Figure 2

Scientific coverage of the SOT and the *Hinode* observatory.

**Figure 3**

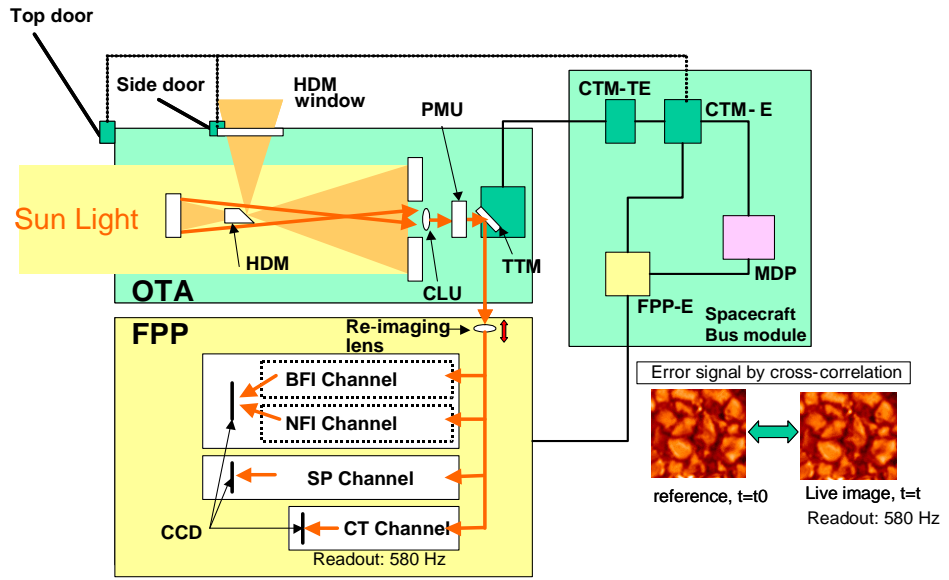

Figure 3

Solar Optical Telescope system overview: OTA: Optical telescope Assembly, FPP; Focal Plane Package, HDM; Heat Dump Mirror, CLU; Collimator Lens Unit, PMU; Polarization Modulation Unit, TTM (referred to as CTM-TM in the text); Tip-Tilt fold Mirror, BFI; Broadband Filter Imager, NFI; Narrowband Filter Imager, SP; Spectro-Polarimeter, CT; Correlation Tracker, MDP; Mission Data Processor, FPP-E; main electronics box for FPP, CTM-E; main electronics box for OTA and TTM, CTM-TE; analog driver for the TTM.

**Figure 4**

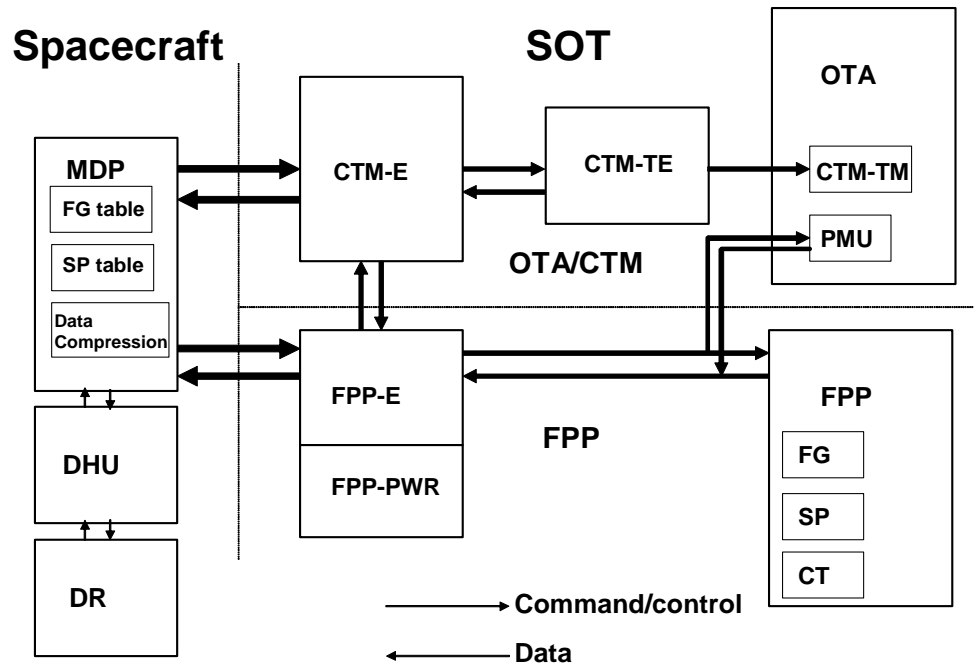

Figure 4

SOT Sub-system instrument configuration: OTA; Optical telescope Assembly, FPP; Focal Plane Package, PMU; Polarization Modulation Unit, CTM-TM; Tip-Tilt fold Mirror, MDP; Mission Data Processor, FPP-E; main electronics box for FPP, FPP-PWR; power supply for FPP subsystem, CTM-E; main electronics box for OTA and TTM, CTM-TE; analog driver for the TTM, DHU; spacecraft Data handling Unit, DR; spacecraft central Data Recorder .

# Optical layout of SOT    Figure 5

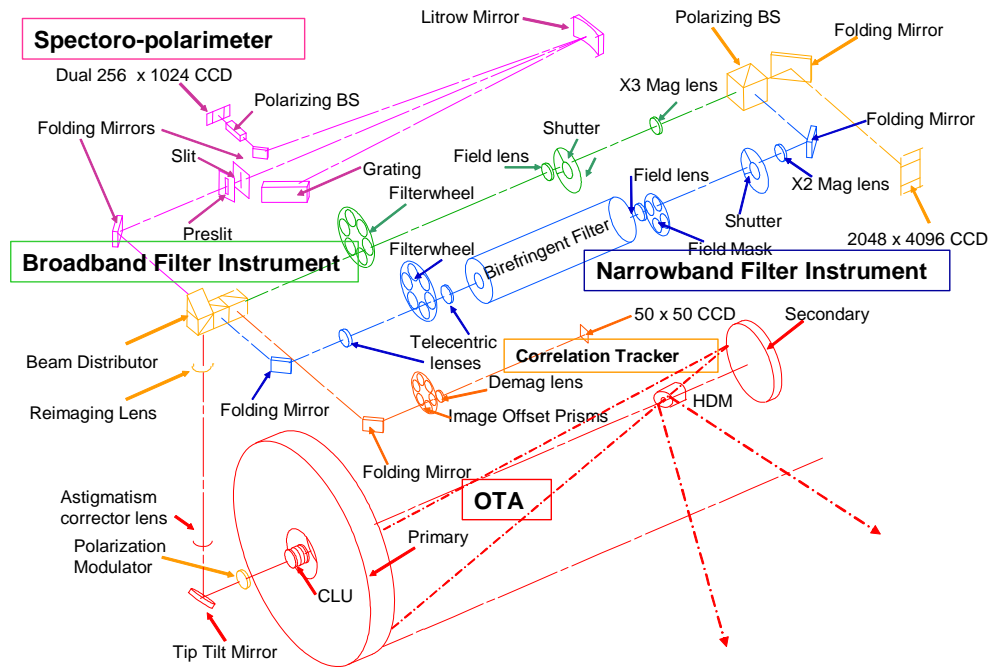

Figure 5

Optical layout of SOT including OTA and FPP

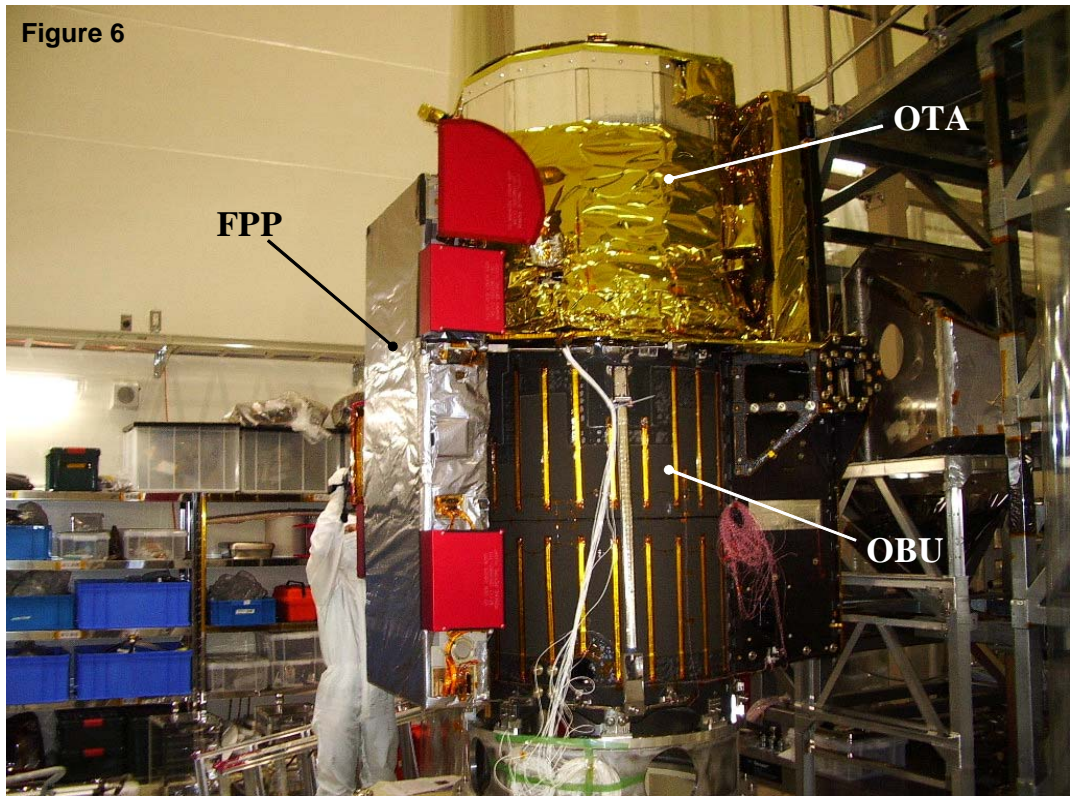

Figure 6

OTA and FPP mounted on the spacecraft optical bench (OBU). The cylindrical optical bench also carries EIS and XRT (not mounted in this photo), and is mounted on the spacecraft bus box. The FPP radiators are covered with red-colored protective covers.

**Figure 7**

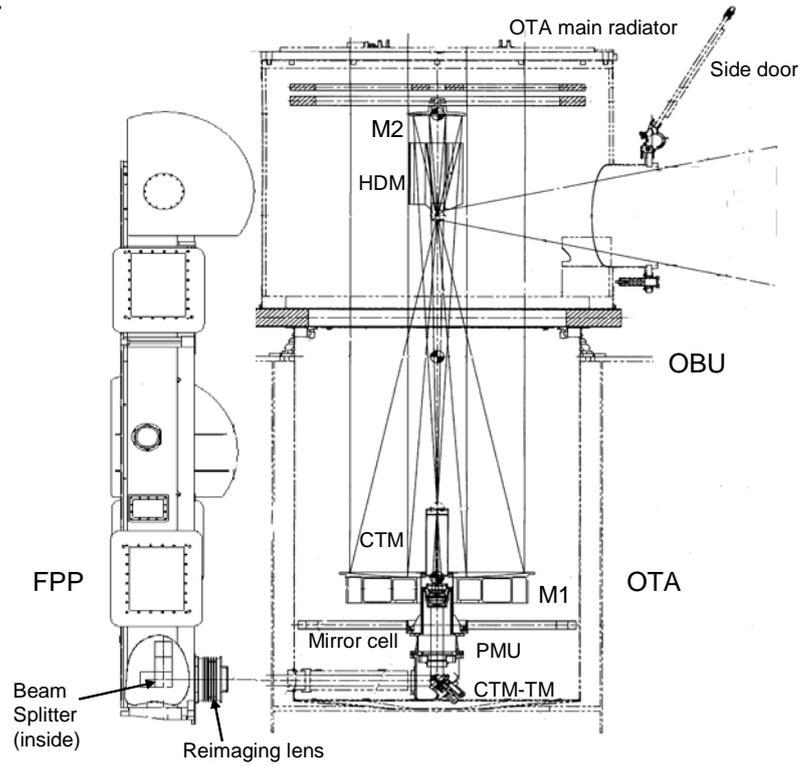

Figure 7

Optical interface between OTA and FPP. OTA and FPP are mounted on the common optical bench unit (OBU). M1; OTA primary mirror, M2; OTA secondary mirror, HDM; Heat Dump Mirror, CLU; Collimator Lens Unit (taken from Suematsu et al 2008).

**Figure 8 (a)**

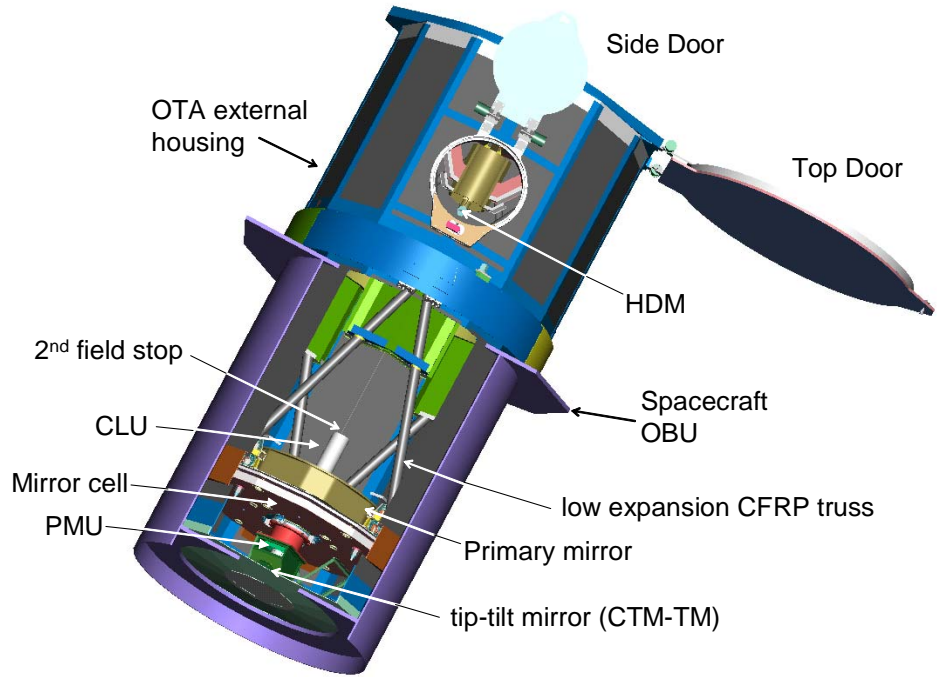

**Figure 8 (b)**

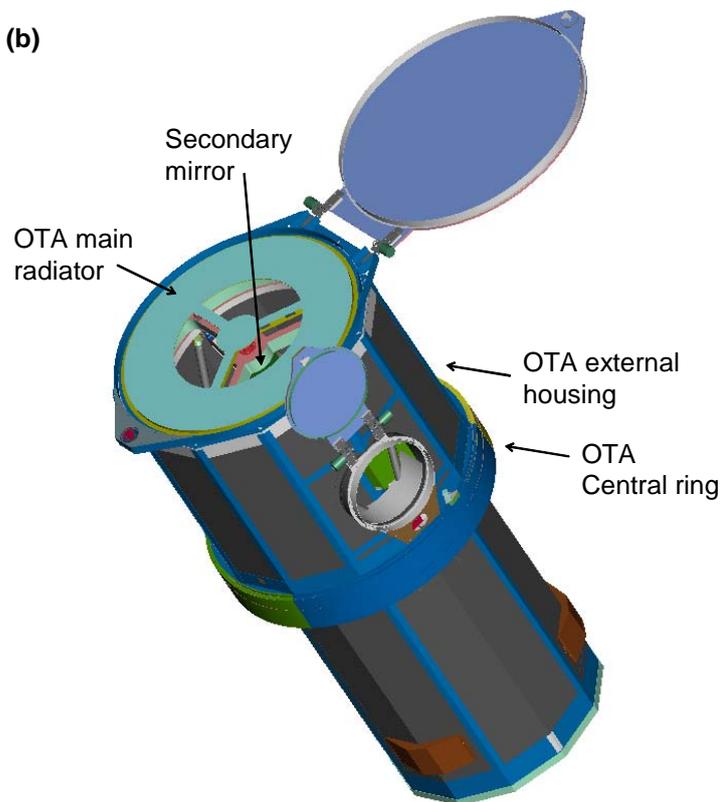

Figure 8 (a) (b)

SOT Optical Telescope Assembly

**Figure 9**

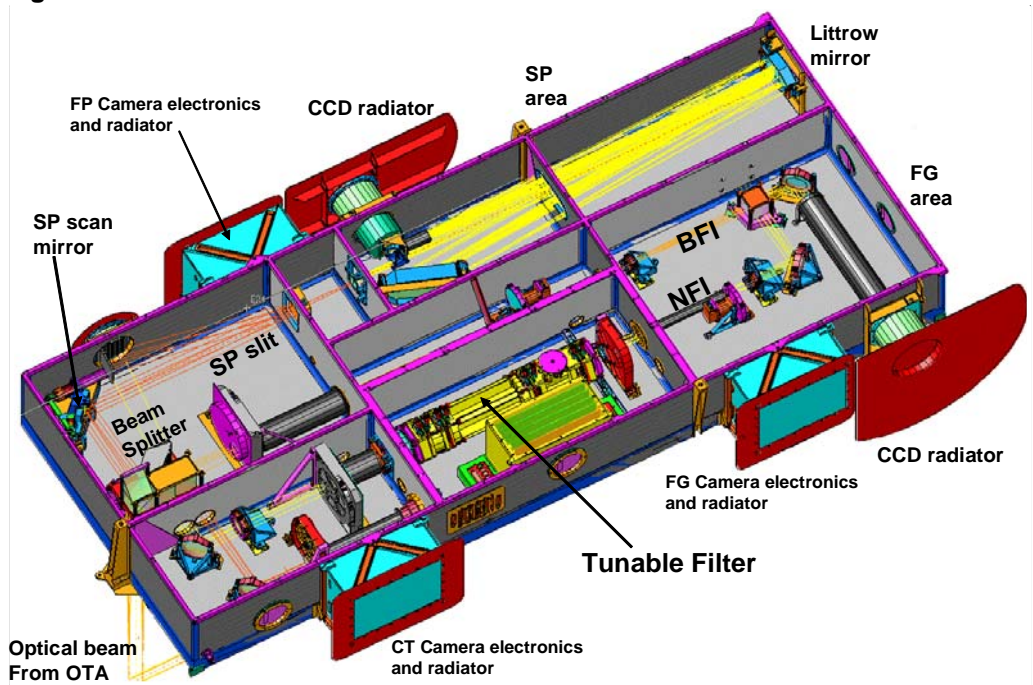

Figure 9

SOT Focal Plane Package (See Tarbell et al., 2008)

**Figure 10**

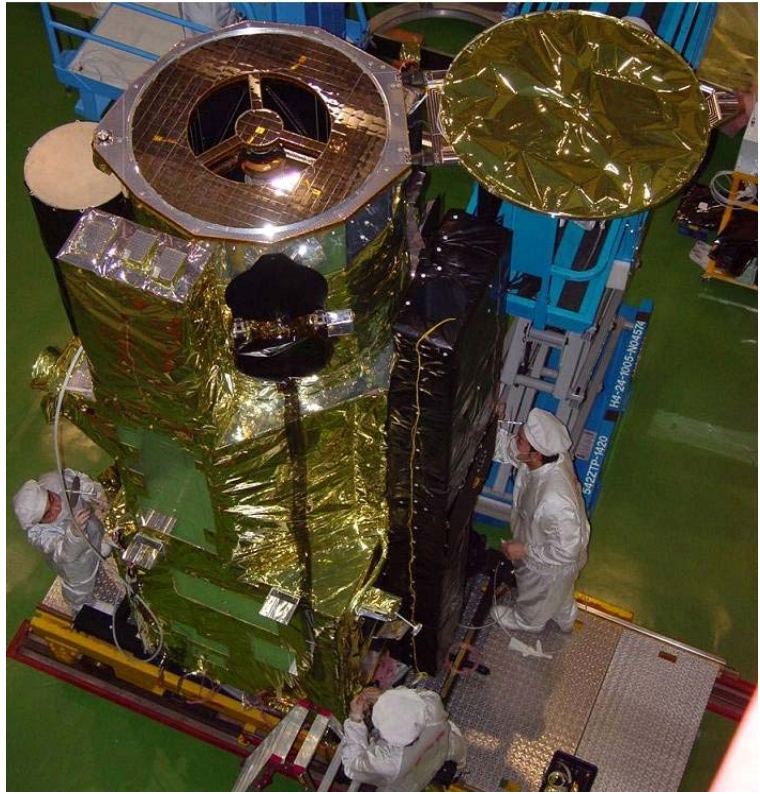

Figure 10

SOT and the *Hinode* satellite for the spacecraft-level thermal vacuum test.

**Figure 11**

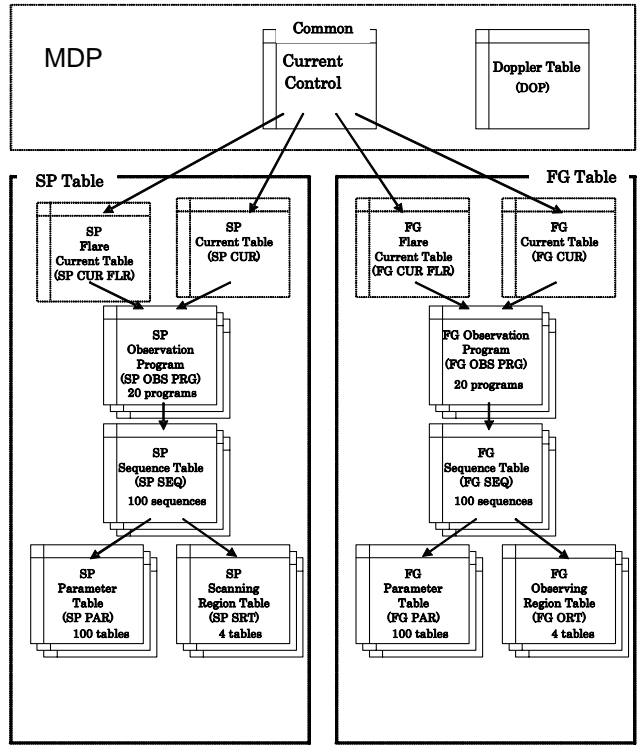

Figure 11

SOT control table structure

**Figure 12**

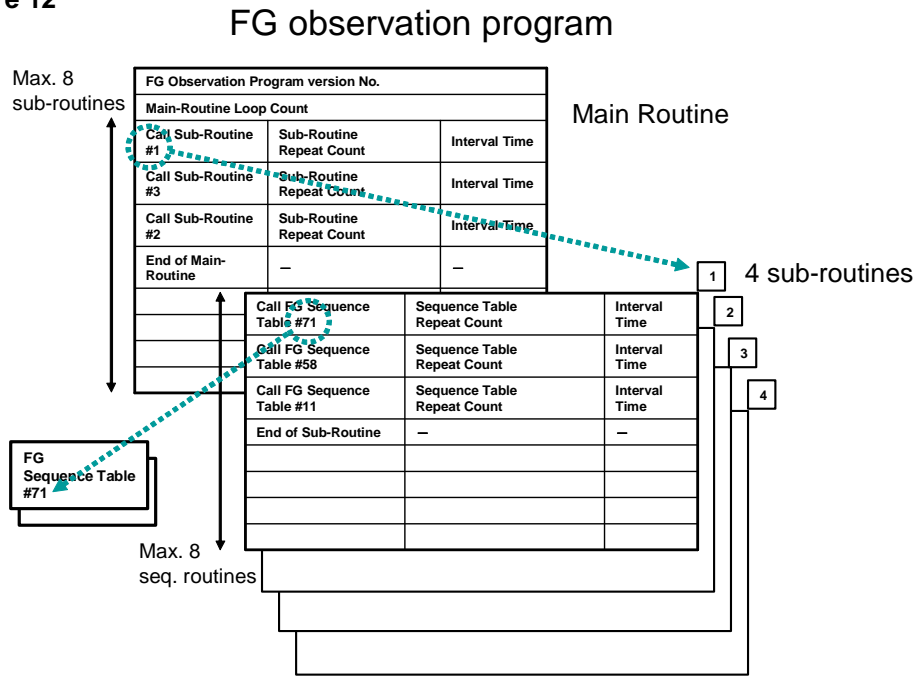

Figure 12

SOT Observation Program structure

Table 1 Optical Telescope Assembly (OTA) overview.

| | | |
|---|---|---|
| Telescope | | |
| | Optics type | Aplanatic Gregorian with heat dump mirror |
| | Primary mirror | 50cm aperture, light-weighted ULE |
| | Primary-secondary mirror length | 1.5 m |
| | Central obscuration ratio | 0.344 in radius |
| | Effective F ratio | 9.055 at secondary focus |
| Collimator lens (CLU) | | |
| | Exit pupil size | 3 cm, collimated in air |
| Polarization modulator (PMU) | | |
| | Rotation speed | Continuous, 1.6sec/rotation |
| Tip-tilt mirror for image stabilizer (CTM) | | |
| | | See Table 5 |

Table 2 SOT/FPP: filter observations.

| Broadband Filter Imager (BFI) | | | | |
|---|---|---|---|---|
| Field of view | | 218×109 arcsec (full FOV) | | |
| CCD | | 4K×2K pixel (full FOV), shared with NFI | | |
| Spatial sampling | | 0.0541 arcsec/pixel (full resolution) | | |
| Spectral coverage | | | | |
| | Center (nm) | Width (nm) | Line of interest | Purpose |
| | 388.35 | 0.7 | CN i | Magnetic network imaging |
| | 396.85 | 0.3 | Ca ii H | Chromospheric heating |
| | 430.50 | 0.8 | Ch i | Magnetic elements |
| | 450.45 | 0.4 | | Blue continuum Temperature |
| | 555.05 | 0.4 | | Green continuum Temperature |
| | 668.40 | 0.4 | | Red continuum Temperature |
| Exposure time | | 0.03−0.8 s (typical) | | |
| Narrowband Filter Imager (NFI) | | | | |
| Field of view | | 328×164 arcsec (unvignetted 264×164 arcsec) | | |
| CCD | | 4K×2K pixel (full FOV), shared with BFI | | |
| Spatial sampling | | 0.08 arcsec/pixel (full resolution) | | |
| Spectral resolution | | 0.009 nm (9pm ) at 630 nm | | |
| Spectral band (tunable filter) | | | | |
| | Center (nm) | Width (nm) | Lines of interest | $g$_eff | Purpose |
| | 517.2 | 0.6 | Mg ib 517.27 | 1.75 | Chromospheric Dopplergrams and magnetograms |
| | 525.0 | 0.6 | Fe i 524.71<br>Fe i 525.02<br>Fe i 525.06 | 2.00<br>3.00<br>1.50 | Photospheric magnetograms |
| | 557.6 | 0.6 | Fe i 557.61 | 0.00 | Photospheric dopplergrams |
| | 589.6 | 0.6 | Na D 589.6 | | Very weak fields (scattering polarization) Chromospheric fields. |
| | 630.2 | 0.6 | Fe i 630.15<br>Fe i 630.2 | 1.67<br>2.5 | Photospheric magnetograms |
| | 656.3 | 0.6 | H i 656.28 | | Chromospheric structure |
| Exposure time | | 0.1−1.6 s (typical) | | |
| Standard observable examples for filter observations | | | | |
| Filtergram | A signal exposure for each spectral coverage | | | |
| | Frame size | 4K×2K, 2K×2K, 1K×2K, or 0.5K×2K | | |
| | Summing | 1×1 (1K×2K or smaller), 2×2, or 4×4 pixel | | |
| | Readout time | 3.4 s (1×1 sum), 1.7 s (2×2), 0.9 s (4×4) Partial readout for faster cadence | | |
| | Reconfigure time | <2.5 s (for changing filter wheels etc) | | |
| Dopplergram | Image of the Doppler shift of a spectral line derived from narrowband filtergrams at several wavelengths | | | |
| | Frame size | 4K×2K, 2K×2K, 1K×2K, or 0.5K×2K | | |
| | Summing | 1×1 (1K×2K or smaller), 2×2, or 4×4 pixel | | |
| | Duration | 12.8 s (4 images, 2×2 sum, 0.8 s exposure) | | |
| Longitudinal magnetogram | Stokes V/I images converted onboard from narrowband filtergrams | | | |
| | Frame size | 2K×1K, 1K×2K, or 2K×2K | | |
| | Summing | 1×1 (1K×2K or smaller), 2×2, or 4×4 pixel | | |
| | Duration | 8 images (4 wavelengths) are taken.<br>12.8 s for 1K×2K and ≈21 s for 2K×2K | | |
| Stokes *IQUV* (for vector magnetogram) | *I/Q/U/V* images made onboard from narrowband filtergrams at different polarization modulator positions | | | |
| | Shuttered exposures | Frame size | 4K×2K, 2K×2K, 1K×2K, or 0.5K×2K | |
| | | Summing | 1×1(1K×2K or smaller), 2×2, or 4×4 pixel | |
| | Shutterless exposures | Frame size | Various | |
| | | Summing | 1×1, 2×2, or 4×4 pixel | |
| | | Duration | 1.6−12.8 s (1−8 waveplate rotations) | |

Table 3 SOT/FPP Spectro-Polarimeter Observations.

| Spectro-Polarimeter (SP) | | | |
|---|---|---|---|
| | Field of view along slit | | 163.84 arcs (NS direction) |
| | Spatial scan range | | 327.62 arcs (transverse to slit, EW direction) |
| | Spatial sampling (slit) | | 0.16 arcsec |
| | Spectral line and coverage | | Fe i 630.15 nm<br>Fe i 630.2 nm<br>Coverage: 630.08 nm to 630.32 nm |
| | Spectral resolution/sampling | | 3pm / 2.15pm |
| | Measurement of polarization | | Stokes *I, Q, U, V* simultaneously with dual beam (orthogonal linear components) |
| | Polarization signal to noise | | $10^3$ (normal map) |
| Standard observable (mapping mode) examples for SP | | | |
| | Normal mapping | Time per position | 4.8 s (3 rotations of waveplate) |
| | | Polarimetric accuracy | 0.001 |
| | | FOV along slit | 164 arcsec |
| | | Sampling along slit | 0.16 arcsec |
| | | Data size | 918K pixels in 4.8 s or 191K pixel s$^{-1}$ |
| | | Slit-scan sampling | 0.16 arcsec |
| | | Time for map area | 50 s for 1.6 arcsec wide<br>83 min for 160 arcsec wide |
| | Fast mapping | Time per position | One rotation for the 1st slit position and another rotation for the 2nd slit position to form one slit data |
| | | FOV along slit | 164 arcsec |
| | | Sampling along slit | 0.32 arcsec |
| | | Data size | 459K pixels in 3.6 s or 127K pixel s$^{-1}$ |
| | | Slit-scan sampling | 0.32 arcsec |
| | | Time for map area | 18 s for 1.6 arcsec wide<br>30 min for 160 arcsec wide |
| | Dynamics | Time per position | 1.6 s (one rotation) |
| | | FOV along slit | 32 arcsec (to reduce data size) |
| | | Sampling along slit | 0.16 arcsec |
| | | Data size | 179K pixels in 1.6 s or 120K pixel s$^{-1}$ |
| | | Slit-scan sampling | 0.16 arcsec |
| | | Time for map area | 18 s for 1.6 arcsec wide |

Table 4 SOT Observation control and data handling.

| | | |
|---|---|---|
| SOT control | Table-driven (Figures 9 and 10) | |
| Effective process speed in MDP | 832K pixels s$^{-1}$ (maximum, for FPP data) | |
| Bit compression in MDP | 16bit data compressed to 12 bit<br>8 lookup tables | |
| Image compression in MDP (expected compression ratio) | 12bit DPCM (lossless) | 6−8bits/pixel |
| | 12bit JPEG(DCT) (lossy) | ≤3bits/pixel for filters<br>≈1.5bits/pixel for SP |
| # Compression rate depends on images and required image quality | | |
| Allocated telemetry rate (max) for SOT | ≈1.3Mbps (nominal)<br>≈1.8Mbps (SOT dominant) | |
| Data rate (after compression) averaged per day | ≈300 kbps, assuming 15 downlink stations in a day | |

Table 5 Image Stabilization System.

| Correlation Tracker (CT) in FPP | | |
|---|---|---|
| | CCD | 50x50 pixels, 0.22 arcsec/pixel |
| | Frame rate | 580Hz |
| | Spectral range | 629−634nm |
| | Displacement Range | ±5 pixels |
| | Error signal accuracy | < 0.01 arcsec |
| | Control | FPP onboard computer |
| Tip-tilt mirror (CTM-TM) in OTA | | |
| | Signal used for closed loop control | Residual signal from correlation tracker |
| | Actuator | 3 commercial Piezo actuators |
| | Tilt range | 10.5 arcsec in radius on the sky |
| | Control crossover frequency | 14 Hz (nominal gain) |
| | Stability | ≈0.007arcsec (one sigma in-orbit) |
| | Control | CTM dedicated computer for servo control (CTEM-E) |